# Observation of Intra- and Inter-band Transitions in the Optical Response of Graphene

**Leandro M. Malard[1], Kin Fai Mak[1], A. H. Castro Neto[2,3], N. M. R. Peres[4], and Tony F. Heinz[1]**

[1]Departments of Physics and Electrical Engineering, Columbia University, 538 West 120th Street, New York, New York 10027, USA
[2]Department of Physics, Boston University, 590 Commonwealth Avenue, Boston, Massachusetts 02215, USA
[3]Graphene Research Centre, National University of Singapore, 2 Science Drive 3, 117542, Singapore
[4]Departamento de Física e Centro de Física, Universidade do Minho, P-4710-057 Braga, Portugal
(April 15, 2011)

The optical conductivity of freely suspended graphene was examined under non-equilibrium conditions using femtosecond pump-probe spectroscopy. We observed a conductivity transient that varied strongly with the electronic temperature, exhibiting a crossover from enhanced to decreased absorbance with increasing pump fluence. The response arises from a combination of bleaching of the inter-band transitions by Pauli blocking and induced absorption from the intra-band transitions of the carriers. The latter dominates at low electronic temperature, but, despite an increase in Drude scattering rate, is overwhelmed by the former at high electronic temperature. The time-evolution of the optical conductivity in all regimes can described in terms of a time-varying electronic temperature.
PACS: 78.67.Wj, 78.47.D-, 78.47.jb

The optical properties of graphene have been the subject of much attention [1-3]. Interest in this topic is generated both by the insight that the optical response of graphene provides into the nature of its excited states and their interactions and by the importance of understanding graphene's optical response for emerging photonic applications [2, 4]. The response of graphene to excitation by femtosecond laser pulses has attracted particular interest [5-11]. Probing dynamics on the femtosecond time scale provides information about electron-electron, electron-phonon, and phonon-phonon interactions and also has important implications for the behavior of recently developed ultrafast photonic devices [2, 4].

To date all ultrafast pump-probe measurements of graphene have observed an induced bleaching of the absorption under optical excitation [7-9]. This behavior arises from Pauli blocking of the strong inter-band optical transitions [7-9] and is expected for electronic temperatures sufficiently high to induced filling of the states. The optical response of graphene arises, however, from two fundamental processes: inter-band and intra-band optical transitions [3, 12-14]. Here we report observation of a *dominant intra-band* contribution to the transient optical response, with a corresponding *enhanced absorption*. We identify this effect through probing freely suspended graphene samples at relatively low pump fluence. With increasing pump fluence, the transient optical response exhibits a crossover from enhanced absorption to bleaching as Pauli blocking of inter-band transitions becomes more prominent.

We explain the observed temporal evolution of the optical absorption for all pump fluences within the framework of a thermalized electronic energy distribution that is in equilibrium with a set of strongly coupled optical phonons. Cooling of this subsystem, and the decrease in electronic temperature, is controlled by energy loss of the strongly coupled optical phonons to lower energy phonons through anharmonic decay. A fluence-independent time constant of 1.4 ps is deduced, which is consistent with recent time-resolved Raman scattering measurements of the cooling of zone-center optical phonons [15]. In addition to the importance of these measurements for understanding ultrafast carrier dynamics and applications in ultrafast photonics, the intra-band response at high carrier temperature provides insight into carrier dynamics in a regime relevant to high-field charge transport in graphene [16-19]. In particular, our experiments reveal a sharp increase in the Drude scattering rate with increasing temperature of the electrons and optical phonons, as also suggested by high-field electrical transport measurements in carbon nanotubes [20, 21] and graphene [16-19]. Our results shown to be compatible with the behavior expected if electron-optical phonon scattering plays a dominant role in determining the carrier scattering rate in this high temperature regime.

In our experimental study we made use of freely suspended graphene samples. Such samples allow us to eliminate both accidental doping effects and energy transfer associated with the substrate. We prepared exfoliated graphene samples that were freely suspended over trenches patterned in transparent $SiO_2$ substrates [22]. The substrates, with trenches of width 4 – 5 μm and depth of ~3 μm (as characterized by atomic force microscopy), were carefully cleaned by chemical etching (nanostrip) before we deposited graphene by mechanical exfoliation from kish graphite. Areas of graphene of single-layer thickness were identified by optical microscopy and further characterized by Raman spectroscopy [22]. The low doping levels of the samples were reflected in the large (~15 cm$^{-1}$) width of the G-mode and characteristic asymmetry of the 2D-mode, while the low level of defects was indicated by the absence of the disorder-induced D-mode.

The optical pump-probe measurements were performed using a modelocked Ti-sapphire laser producing pulses of ~300-fs duration at an 80-MHz repetition rate. The 800-nm wavelength output of the laser provided the probe pulses, while radiation at 400 nm, obtained by frequency doubling in a β-barium borate (BBO) crystal, served as the pump excitation. Both the pump and probe beams were focused onto the samples with a single 40× objective to yield Gaussian spots of comparable widths of ~1.5 μm (FWHM). The absorbed pump fluence was determined using the absorbance of graphene at 400 nm (1.8πα = 4.14 % [23]). To account for the spatial variation of the pump and probe beams, the effective fluence was determined by weighting the absorbed probe fluence using the spatial profile of the probe beam. Effective absorbed fluences $F$ between 0.5 and 4.0 $\mu J/cm^2$ were investigated experimentally. We note that over this range of pump fluences, saturation of absorbance in graphene at the pump wavelength is negligible [4]. The pump-probe measurements were performed by modulating the pump laser at 20 kHz and detecting the synchronous change in probe transmission. The induced modulation of the probe beam for the lowest pump fluence was ~$10^{-6}$. For a suspended thin film of material like our graphene sample, the fractional change in transmittance, $\Delta T/T$, is given by the change in the sample absorbance $\Delta A$. The later is proportional to the change in the real part of the optical sheet conductivity of graphene, $\sigma^{(1)}$, according to $\Delta A = (4\pi/c)\Delta\sigma^{(1)}$ [14]. We can therefore directly convert our experimentally observed change in transmission into a fundamental material property of the graphene, namely, $\Delta\sigma^{(1)}$.

Figure 1(a) shows the measured transient response of graphene for a comparatively low absorbed pump fluence of $F = 0.5\ \mu J/cm^2$. An increase in the optical conductivity (enhanced absorption) is seen. The rise time of this response is comparable to our experimental time resolution, while the relaxation can be fit to a single exponential decay with a time constant of $\tau_{exp} = 3.1$ ps.

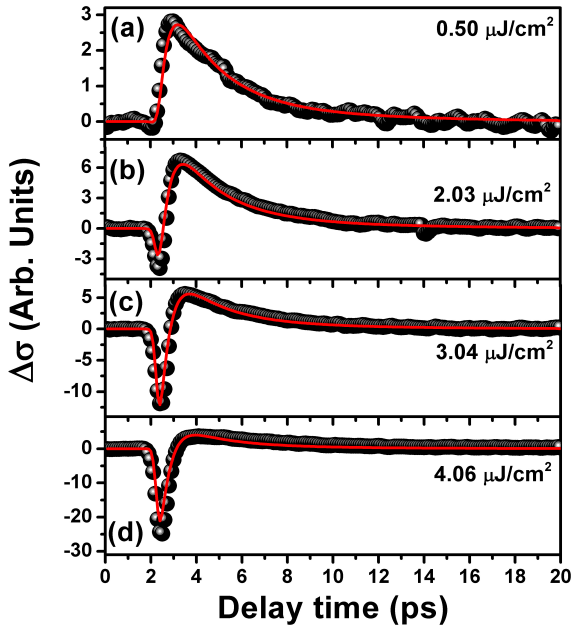

FIG. 1 (color online): Measured dynamics (closed circles) of the transient optical conductivity of graphene for excitation with different absorbed pump fluences $F$: (a) 0.50 $\mu J/cm^2$, (b) 2.03 $\mu J/cm^2$, (c) 3.04 $\mu J/cm^2$, and (d) 4.06 $\mu J/cm^2$. The red curves are fits based on the model described in the text, which include both intra- and inter-band contributions to the optical response.

To analyze our data, we consider models for the optical conductivity of graphene with the electronic system described by a Fermi-Dirac distribution at (electronic) temperature $T$. The assumption of a thermalized energy distribution for the electronic excitations is justified by the time resolution of >100 fs in our measurements. On this time scale, the excited electrons and holes are equilibrated with one another, as well with the strongly coupled optical phonons (near the $\Gamma$- and $K$-points) [5, 24].

The predicted change in the optical conductivity of graphene under excitation arises from both intra- and inter-band contributions, $\Delta\sigma^{(1)} = \Delta\sigma^{(1)}_{intra} + \Delta\sigma^{(1)}_{inter}$. For the case of thermalized electronic energy distributions relevant in our case, the optical response of graphene has been examined in several theoretical investigations [3, 12, 13]. Within a picture of non-interacting electrons, the induced change in the real part of the optical sheet conductivity follows directly from the corresponding changes in the intra- and inter-band terms:

$$\sigma^{(1)}_{intra} = \frac{8\ln 2}{\pi}\frac{\pi e^2}{2h}\frac{\Gamma k_B T}{(\hbar\omega)^2 + \Gamma^2}$$

$$\sigma^{(1)}_{inter} = \frac{\pi e^2}{2h}\tanh\left(\frac{\hbar\omega}{4k_B T}\right) \qquad (1)$$

Here $\hbar\omega$ denotes the photon energy of the probe beam, and $\Gamma$ is the carrier scattering rate. For our suspended graphene samples, we have negligible doping and take the chemical potential to be at the Dirac point. We also neglect any transient separation of the chemical potentials for the electrons and holes. While a dynamical separation of the chemical potentials is common in semiconductors under ultrafast excitation, on the relevant time scale the effect is expected to be insignificant in graphene because of the presence of rapid Auger processes for this zero-gap material [4, 11].

The photo-induced change in the optical conductivity in Eqn. (1) from intra-band optical transitions, $\Delta\sigma^{(1)}_{intra}$, is positive in sign. The enhanced conductivity arises from the presence of additional free carriers after optical excitation and scales linearly with the electronic temperature $T$. The inter-band term $\Delta\sigma^{(1)}_{inter}$ yields a negative contribution to the conductivity, a bleaching of the absorption, from Pauli blocking of the strong inter-band transitions at the photon energy $\hbar\omega$. In our regime of $\hbar\omega = 1.55\ eV \gg k_B T$, the fractional change in the inter-band conductivity is slight and varies nearly exponentially in $T$. Accordingly, the intra-band contribution dominates at comparatively low temperatures (low fluence), but, as shown in Fig. 2, is overwhelmed by the inter-band terms as the temperature increases. Thus the transient *increase* in absorption seen in Fig. 1(a) for low pump fluence reflects the change in *intra-band* optical response.

In order to apply the analysis of graphene optical response presented in Eqn. (1) to our experiment, we introduce a simple model to describe the temporal evolution of the electron temperature $T(t)$ induced by the pump laser. As we shall see, this treatment predicts transient optical conductivities compatible both with the low-fluence data already presented and the high-fluence response discussed below. To describe $T(t)$, we first note that on the time scale of our measurement, the deposited energy from the pump pulses rapidly equilibrates among the electronic excitations and the strongly coupled optical phonons [5, 24]. Energy then leaves this coupled subsystem through anharmonic decay of the phonons on the picosecond time scale, which we describe phenomenologically by an effective phonon lifetime $\tau_{ph}$ [25]. To determine the resulting electronic temperature $T(t)$ we need to know the heat capacity of the sub-system consisting of the electrons and the strongly coupled phonons. For this purpose, we use an Einstein model consisting of two phonon modes, the Γ- and K-point phonons. Since only a small part of the phonon branch gets populated, we considered only a fraction of the graphene Brillouin zone to obtain the phonon heat capacity. The fraction is determined by comparing results from [24] and the initial condition is fixed by the absorbed fluence $F$. (For details of this analysis, please refer to the Auxiliary Materials [26].) The electronic contribution to the heat capacity is minor and is neglected. Using this heat capacity, we the find $T(t)$ for any given excitation condition and, using Eqn. (1), then obtain the temporal evolution of the optical conductivity. In fitting our results, we treating the phonon lifetime $\tau_{ph}$ and the Drude scattering rate Γ as adjustable parameters. As we discuss below, we use an effective scattering rate Γ that depends on the pump fluence, but is independent of time, to account for the variation of the scattering rate with the temperature of the electrons and optical phonons.

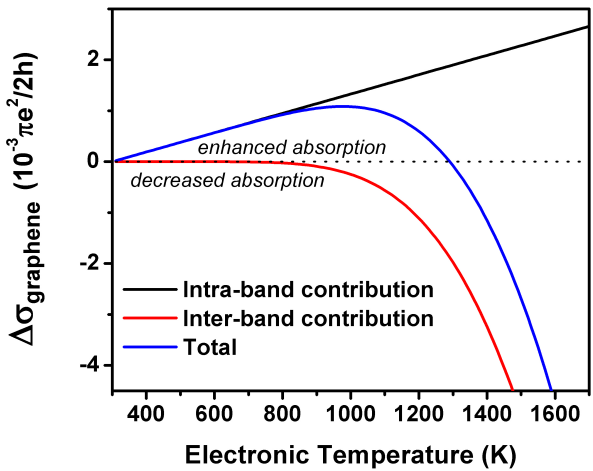

Fig. 2 (color online): Change in the optical conductivity of graphene as a function of electronic temperature calculated from Eq. (1). Black and red curves show the expected contributions from intra- and inter-band terms, respectively. The blue curve shows the total response.

The measured time evolution of the optical conductivity in Fig. 1(a) is reproduced well with this model. We use an effective scattering rate of Γ = 27 meV and a phonon lifetime of $\tau_{ph}$ =1.4 ps. This lifetime differs significantly from time constant ($\tau_{exp}$ = 3.1 ps) of the conductivity transient, a difference that reflects the strong temperature dependence of the phonon heat capacity: The electronic temperature, which tracks that of the strongly coupled optical phonons and is described by $\tau_{exp}$, falls more slowly than the energy content of the optical phonons, which is characterized by $\tau_{ph}$. The inferred phonon lifetime of $\tau_{ph}$ =1.4 ps, we note, lies between that obtained by time-resolved Raman scattering measurements for graphite (2.2 ps) [24] and for graphene (1.2 ps) [15]. The slightly longer phonon lifetime compared to that measured for graphene [15] is attributed to the fact that our graphene samples are suspended, thus eliminating possible decay channels involving the generation with surface polar phonons in the substrate [16, 17, 19].

At higher pump fluences (and, correspondingly, higher electronic temperatures), the optical conductivity is expected to decrease as state filling of the inter-band transitions begins to dominate the transient response (Fig. 2). The measured conductivity transients at higher pump fluence are shown in Fig. 1 (b-d). With increasing pump fluence, the initial response is indeed negative. At later times when the electronic temperature drops, the intra-band contribution becomes dominant and a positive conductivity change is observed.

The experimental data of Fig. 1(b-d) can be fit (red curves) with the same model of the electronic temperature described above, but now with the inclusion of the inter-band optical response of graphene. Thus model can then accurately describe the dynamics of the conductivity transients at all fluences [Fig. 1(a-d)]. In this fitting process, we use a single, fluence-independent parameter of $\tau_{ph}$ =1.4 ps for the optical phonon lifetime. This parameter reflects the anharmonic decay rate of these optical phonons. It should not increase with pump fluence unless the excitation of the resulting phonon decay modes also increased significantly [27], which is not expected this regime. On the other hand, the fitting procedure implies that the effective carrier scattering rate Γ increases with pump fluence. The inferred variation of Γ is shown in Fig. 3 as a function of absorbed fluence (bottom scale) and of the peak electronic temperature (top scale) in our model. At high fluence, corresponding to a peak temperature of 1,700 K, we find Γ = 52 meV. Both this rate and that inferred for lower fluences are significantly higher than the scattering rates implied by conventional transport measurements [28] obtained for a more modest temperature range.

To understand the origin of the enhanced scattering rate, recall that the electron-phonon coupling in graphene is especially strong for the optical phonon modes located at the center and edge of the Brillouin zone [5, 25]. This coupling is not, however, important for conventional low-bias charge transport measurements at room temperature. In this conventional regime, the optical phonons are not thermally excited and cannot be absorbed in a scattering event, while the electron energy is too low to scatter through the emission of optical phonons. This situation changes when the temperature of the electrons and optical phonons becomes comparable to that characterizing the optical phonon, $\hbar\omega / k_B$ [20, 21]. Electron scattering with optical phonons is then allowed and the contribution of this scattering process to the temperature dependence of the Drude is significant [29].

To obtain specific predictions for the increased Drude scattering rate Γ from electron – optical phonon interactions, we consider the zone-center and zone-edge phonons to be dispersionless branches with energies of 200 meV and 150 meV, respectively. We then can then calculate (see Auxiliary Materials [26]), the expected temperature dependence of Γ from electron- optical phonon scattering. As shown in Fig. 3, this contribution increases nearly linearly with $T$ for the relevant temperatures. By using published values for the electron-phonon coupling for both the zone-center [30] and zone-edge [31] optical phonons, we obtain semi-quantitative agreement with the experimental results for Γ as a function of pump fluence. Better agreement is not anticipated, since the treatment involves several significant simplifications. We first note that the deduced effective Drude scattering rate Γ actually corresponds to fitting the response over a range of electronic temperatures. Here we assume that the result is dominated by the behavior near the peak electronic temperature. More elaborate modeling could take this effect into account. In terms of the underlying description of the optical conductivity at high temperature, our treatment could be improved by a more detailed analysis of the different phonon modes, considered here as two dispersionless branches, involved in the scattering process. More fundamentally, one should also examine possible contributions to $\Gamma$ from electron-electron scattering processes [32], which are known to influence the dc conductivity [33]. Finally, at high electronic temperatures ($T > 1600$K), the experimental values for $\Gamma$ level off with increasing $T$. This effect is absent in the model and suggests that screening of the electron interactions [34] may play a significant role in this regime of high carrier densities.

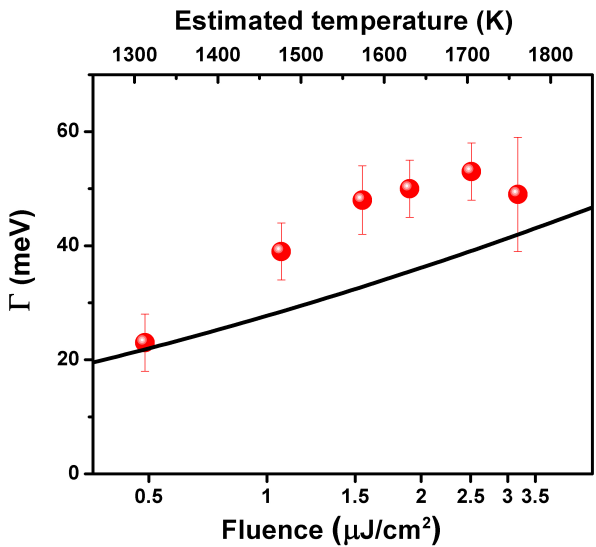

Fig. 3 (color online): Drude scattering rate $\Gamma$ (closed circles) as a function of absorbed pump fluence (bottom scale) and the inferred peak electronic temperature (top scale). The points are inferred from modeling of the experimental transient optical response of the graphene sample. The black line corresponds to the predicted temperature dependence of $\Gamma$ arising from electron- optical phonon interactions, as described in the text and in the Auxiliary Materials [26].

An interesting point of comparison for our results is with those obtained in high-field dc transport studies. We find that the values for $\Gamma$ deduced here for high electronic temperatures are also broadly consistent with those obtained in high-field transport measurements in metallic carbon nanotubes at current saturation [20, 21]. In this regime, electron- optical phonon scattering is also considered to be the dominant contribution to $\Gamma$. For high-bias transport measurements in graphene [16, 17, 19], the carrier scattering rate is also significantly enhanced compared to the low-field behavior. This effect has, however, been attributed to interaction with of the carriers with polar phonons in the substrate on which the graphene is deposited. Our results imply that even for suspended graphene, which lacks these substrate-mediated interactions, the scattering rate will be strongly enhanced at elevated electronic temperatures. This effect will places a fundamental limit on current flow in graphene at electrical high bias.

In conclusion, we have examined ultrafast carrier dynamics in freely suspended graphene samples by optical pump-probe measurements. A crossover of the graphene optical conductivity transients from enhanced to decreased absorption is observed as the pump fluence is increased. This behavior can be understood as the result of the existence of both intra- and inter-band contributions to the optical response of graphene. Analysis of the data implies an optical phonon lifetime of 1.4 ps and enhanced carrier scattering rates at high electronic temperatures. Our experiment demonstrates the importance of free carrier absorption in the visible spectral range for graphene under non-equilibrium conditions and opens up new opportunities for probing fundamental charge transport properties of this 2-dimensional system by means of optical measurements.

We thank Drs. Y. Wu, M. Koshino, D. Song and H. Yan for fruitful discussions and C. H. Lui for sample preparation. We acknowledge support from the Nanoscale Science and Engineering Initiative of the National Science Foundation and from the MURI program of the Air Force Office of Scientific Research. AHCN acknowledges the partial support of the U.S. DOE under grant DE-FG02-08ER46512, and ONR grant MURI N00014-09-1-1063. LM acknowledges support from the CNPq program in Brazil.

# Auxiliary Material

# Observation of Intra- and Inter-band Transitions in the Optical Response of Graphene

**Leandro M. Malard[1], Kin Fai Mak[1], A. H. Castro Neto[2,3], N. M. R. Peres[4], and Tony F. Heinz[1]**

[1]*Departments of Physics and Electrical Engineering, Columbia University, 538 West 120th Street, New York, New York 10027, USA*

[2]*Department of Physics, Boston University, 590 Commonwealth Avenue, Boston, Massachusetts 02215, USA*

[3]*Graphene Research Centre, National University of Singapore, 2 Science Drive 3, 117542, Singapore*

[4]*Departamento de Física e Centro de Física, Universidade do Minho, P-4710-057 Braga, Portugal*

**1) Determination of the electronic temperature**

Here we describe the model employed for the heat capacity of the strongly coupled optical phonons. These excitations are assumed to be in thermal equilibrium with the electronic excitations, but, because of the very low heat capacity of electronic excitations in graphene, the heat capacity is dominated by the phonons. Knowledge of this heat capacity then allows us to convert the experimental value of the absorbed pump fluence into the electronic temperature of this subsystem. With the temperature in hand, we predict the optical response from Eq. (1) of the main text to calculate for a comparison with the experimental data.

Strong electronic coupling is present for the zone-center phonons (G band, energy of 200 meV) and the zone-edge phonons (D band, energy of 150 med) [A1, A2, A3]. Using the Einstein model for heat capacity including these two phonon modes, we obtain for the energy density (per unit area) as a function of temperature [A4]:

$$u(T) = \frac{A}{(2\pi)^2} f \left[ \frac{\hbar\omega_G}{e^{\frac{\hbar\omega_G}{k_B T}} - 1} + \frac{\hbar\omega_D}{e^{\frac{\hbar\omega_D}{k_B T}} - 1} \right]. \quad (E1)$$

Here A is the Brillouin zone area, $f$ is the fraction of Brillouin zone filled by the G and D optical phonons and $\hbar\omega_{G,D}$ denotes the G or D phonon energy. We determined $f$ in Eq. (E1) by comparison with a previous experiment in which transient phonon temperatures for femtosecond laser excitation of graphite were determined by time-resolved Raman scattering [A5, A6]. Fitting these earlier measurements to a polynomial yields the following relation between the absorbed fluence (or energy density) per graphene layer and the temperature:

$$F(T) = 1.76 \times 10^{-7} + 1.605 \times 10^{-19}(-4.79 \times 10^{-9} T + 4.55 \times 10^{6} T^2 \\ + 1484 T^3 + 0.3225 T^4) \ \mu J/cm^2 \quad (E2)$$

We find the fraction $f$ by comparing Eq. (E1) and (E2). Eq. (E1) then yields the energy density as a function of temperature, even for temperatures below the range for which the experimental data from the earlier experiments are available.

To model the temporal evolution of the sub-system of the electronic excitations and the strongly coupled phonons, we assume that the energy relaxes as a rate of $1/\tau_{ph}$ through the anharmonic decay, with Eq. (E1) yielding the variation of the temperature. This temperature dynamics is then insert in Eq. (1) from the main text of the paper to fit our pump-probe dynamics experiment. Because of the non-linear relationship between the energy density and temperature, the decay time of these two quantities are not the same, as can see in Fig. S1.

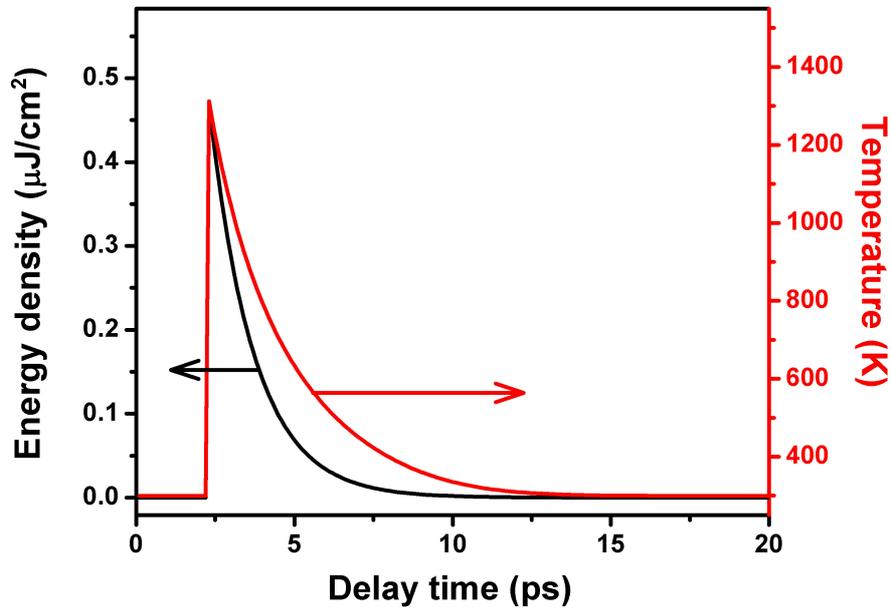

Figure S1: (color online) Temporal evolution of the energy density (black) and temperature (red) calculated for the case of an absorbed fluence of 0.5 $\mu J/cm^2$. The decay times for the energy density and temperature are, respectively, 1.4 and 3.1 ps.

**2) Calculation of the electronic scattering rate in graphene at high temperatures**

The intra-band contribution to the optical conductivity scales with the electron scattering rate Γ. For the regime of temperature of the electrons and strongly coupled optical phonons, we expect that scattering of the electrons by these phonons will contribute significantly to the overall scattering rate Γ. Here we calculate this contributions and find that it yields a rate similar to that inferred from the experiment. As above, we model the strongly coupled phonons by two branches of dispersionless optical phonons, the 200 meV zone-center and 150 meV zone-edge phonons.

The self-energy of the electrons in graphene due to optical phonons is given by [A7]:

$$\Sigma(\vec{k},i\omega_n)/\hbar = -\frac{9}{2}\left(\frac{\partial t}{\partial a}\right)^2 \frac{\hbar}{M_c\hbar\omega_0}\frac{1}{N_c}\sum_{\vec{Q}}\frac{1}{\beta\hbar}\sum_m D^0(i\nu_m)G^0_{bb}(\vec{k}-\vec{Q},i\omega_n-i\nu_m), \text{ [E3]}$$

where $(\partial t/\partial a)$ is the electron-phonon coupling as defined in Ref. [A8], $a$ is the carbon-carbon distance, $t$ is the hopping in the tight-binding model for graphene, $M_C$ is the carbon mass, $\omega_0$ is the phonon frequency. All the other parameters are equally defined by Ref. [A7]. After performing the Matsubara summation, the imaginary part of the self-energy is given by

$$\Im\Sigma(\vec{k},i\omega_n)/\hbar = -\frac{9\pi}{4}\left(\frac{\partial t}{\partial a}\right)^2 \frac{\hbar}{M_c\hbar\omega_0}\frac{1}{N_c}\sum_{\vec{Q}}$$
$$+\left[n_B(\omega_0)+n_F(\varepsilon_{\vec{k}-\vec{Q}})\right]\delta(\varepsilon_{\vec{k}}-\varepsilon_{\vec{k}-\vec{Q}}+\hbar\omega_0)$$
$$+\left[n_B(\omega_0)+1-n_F(\varepsilon_{\vec{k}-\vec{Q}})\right]\delta(\varepsilon_{\vec{k}}-\varepsilon_{\vec{k}-\vec{Q}}-\hbar\omega_0) \quad \text{[E3]}$$
$$+\left[n_B(\omega_0)+n_F(-\varepsilon_{\vec{k}-\vec{Q}})\right]\delta(\varepsilon_{\vec{k}}+\varepsilon_{\vec{k}-\vec{Q}}+\hbar\omega_0)$$
$$+\left[n_B(\omega_0)+1-n_F(-\varepsilon_{\vec{k}-\vec{Q}})\right]\delta(\varepsilon_{\vec{k}}+\varepsilon_{\vec{k}-\vec{Q}}-\hbar\omega_0),$$

where $n_B$ and $n_F$ are the Bose and Fermi distribution functions, $\varepsilon_{\vec{k}} = v_F\hbar k$ and $v_F = 3ta/2\hbar$. The two first terms are intra-band scattering processes and the last two are inter-band ones.

The scattering rate is obtained from

$$\tau_{\vec{k}}^{-1} = -\Im\Sigma(\vec{k},\varepsilon_{\vec{k}})/\hbar. \text{ [E4]}$$

To compute $\tau_{\vec{k}}^{-1}$, we need to solve the following integrals appearing in Eq. [E3]:

$$I_\pm = \frac{1}{N_c} \sum_{\vec{Q}} \delta(\varepsilon_{\vec{k}} - \varepsilon_{\vec{k}-\vec{Q}} \pm \hbar\omega_0)$$

$$= \frac{a_c}{(2\pi)^2} \int_0^{2\pi} d\theta \int_0^Q Q dQ \delta(\varepsilon_{\vec{k}} - \sqrt{\varepsilon_{\vec{k}}^2 + \varepsilon_{\vec{Q}}^2 - 2\varepsilon_{\vec{Q}}\varepsilon_{\vec{k}}\cos\theta} \pm \hbar\omega_0) \quad [E5]$$

$$J_- = \frac{a_c}{(2\pi)^2} \int_0^{2\pi} d\theta \int_0^Q Q dQ \delta(\varepsilon_{\vec{k}} + \sqrt{\varepsilon_{\vec{k}}^2 + \varepsilon_{\vec{Q}}^2 - 2\varepsilon_{\vec{Q}}\varepsilon_{\vec{k}}\cos\theta} - \hbar\omega_0),$$

where $a_c = 3\sqrt{3}a^2/2$. The $\delta$-functions in Eq. [E3] guarantee that we can replace $\varepsilon_{\vec{k}-\vec{Q}}$ in the Fermi functions with $\varepsilon_{\vec{k}} \pm \hbar\omega_0$ for the first and second integral, and with $\hbar\omega_0 - \varepsilon_{\vec{k}}$ in the fourth one. The third integral does not contribute to the scattering rate, since its argument is always positive. Computing the three integrals, we find that the scattering rate of an electron with momentum $\vec{k}$ has, as a consequence of phase-space restrictions, different forms for different energy ranges:

1. For $\varepsilon < \hbar\omega_0$:

$$\tau_\varepsilon^{-1} = C\left[2n_B(\hbar\omega_0)\hbar\omega_0 + (\hbar\omega_0 - \varepsilon)\right]$$
$$+ C\hbar\omega_0[n_F(\varepsilon + \hbar\omega_0) - n_F(\varepsilon - \hbar\omega_0)] \quad [E6]$$
$$+ C\varepsilon[n_F(\varepsilon + \hbar\omega_0) + n_F(\varepsilon - \hbar\omega_0)];$$

2. For $\varepsilon > \hbar\omega_0$:

$$\tau_\varepsilon^{-1} = C\left[2n_B(\hbar\omega_0)\hbar\omega_0 + (\varepsilon - \hbar\omega_0)\right]$$
$$+ C\hbar\omega_0[n_F(\varepsilon + \hbar\omega_0) + n_F(\varepsilon - \hbar\omega_0)] \quad [E7]$$
$$+ C\varepsilon[n_F(\varepsilon + \hbar\omega_0) - n_F(\varepsilon - \hbar\omega_0)],$$

where C is given by

$$C = \frac{3\sqrt{3}}{2}\left(\frac{\partial \ln t}{\partial a}\right)^2 \frac{\hbar}{M_c \hbar\omega_0} \quad [E8]$$

and $M_c = 3.2 \times 10^{-27}$ eV·s$^2$/Å. The electron phonon couplings $(\partial t/\partial a)$ used for our calculations are given by $(\partial t/\partial a)_{200 meV} \approx 6.4$ eV/Å [A8] and $(\partial t/\partial a)_{150 meV} \approx 14$ eV/Å [A9] for the zone-center and zone-edge phonons, respectively.

Now we need to average the energy dependent scattering rate given by Eq. [E6-E7] to find the average scattering rate [A10]:

$$<\tau_\varepsilon^{-1}> = \frac{\int d\varepsilon D(\varepsilon) n_F(\varepsilon,T) \tau_\varepsilon^{-1}}{\int d\varepsilon D(\varepsilon) n_F(\varepsilon,T)}, \text{[E9]}$$

where $D(\varepsilon)$ is the density of states of graphene.

Finally to compute the results for the temperature dependence of the scattering rate $\Gamma$ shown in Fig. 3 of the main manuscript, we summed the contributions from the two optical phonon branches:

$$\Gamma \equiv <\tau_\varepsilon^{-1}>_{200 meV} + <\tau_\varepsilon^{-1}>_{150 meV}. \text{[E10]}$$